\documentstyle[aps,floats]{revtex}

\begin{document}

\title{\hbox to \hsize{\hfill\vbox{ 
\hbox{MADPH-01-1240} 
\hbox{AMES-HET-01-08} 
\hbox{hep-ph/0108090}}}
\vspace*{.2in}
Neutrino Superbeam Scenarios at the Peak\footnote{Snowmass 2001 Workshop Contribution}}
\author{V. Barger and D. Marfatia}
\address{Department of Physics, University of Wisconsin,
Madison, WI 53706, USA}
\author{K. Whisnant}
\address{Department of Physics and Astronomy, Iowa State University,
Ames, IA 50011, USA}


\maketitle

\begin{abstract}
We discuss options for U.S. long baseline neutrino experiments
using upgraded conventional neutrino beams, assuming $L/E_\nu$ is chosen
to be near the peak of the leading oscillation. We find that 
for $L = 1290$ km (FNAL--Homestake) or 1770~km (FNAL--Carlsbad, or
BNL--Soudan) it is possible to simultaneously have
good $\sin^22\theta_{13}$ reach and sgn($\delta
m^2_{31}$) determination, and possibly sizeable $\tau$ rates and
some $\delta$ sensitivity.
\end{abstract}

\medskip

In this report we discuss possible three--neutrino scenarios for long
baseline neutrino experiments using upgraded conventional neutrino beams
(superbeams). In each case we examine their ability to measure
$\nu_\mu \to \nu_e$ and $\nu_\mu \to \nu_\tau$ appearance, discover $CP$
violation, and to determine the sign of the leading $\delta
m^2$. Details of our calculations can be found in Ref.~\cite{bmw}. For
the $\nu_\mu \to \nu_e$ oscillation probability we use the approximate
analytic expressions of Ref.~\cite{cervera,freund}, which are particularly
helpful in determining the general properties described below. We
emphasize that many other beam design and source--detector
configurations are possible; the scenarios discussed here illustrate
some of the capabilities of such facilities.

We choose five distances 
that could be appropriate for likely
proton driver and detector sites (see Table~\ref{tab:baselines}): 
350~km (BNL--Cornell, or similar to the
295~km of JHF--SK), 730~km (FNAL--Soudan or CERN--Gran Sasso), 1290~km
(FNAL--Homestake, or similar to the 1200~km of JHF--Seoul), 1770~km
(FNAL--Carlsbad, or similar to the 1720~km of BNL--Soudan), and 2900~km
(FNAL--SLAC, or similar to the 2920~km of BNL--Carlsbad). The latter
distance would also be similar to FNAL--San Jacinto (2640~km) or
BNL--Homestake (2540~km).
\begin{table}[t]
\squeezetable
\caption[]{Baseline distances for some detector sites (shown in
parentheses) for neutrino beams from FNAL, BNL, JHF, and CERN.}
\label{tab:baselines}
\begin{tabular}{llll}
\multicolumn{4}{c}{Beam source}\\
FNAL & BNL & JHF & CERN\\
\hline
\hline
& \phantom{1}350 (Cornell) & \phantom{1}295 (Super--K) & \\
\hline
\phantom{1}730 (Soudan) & & & \phantom{1}730 (Gran Sasso)\\
\hline
1290 (Homestake) & & 1200 (Seoul) & \\
\hline
1770 (Carlsbad) & 1720 (Soudan) & & \\
\hline
2640 (San Jacinto) & 2540 (Homestake) & & \\
2900 (SLAC) & 2920 (Carlsbad) & & \\
\end{tabular}
\end{table}

For each $L$, we choose $\langle E_\nu \rangle$ such that $\Delta = 1.27
\delta m^2_{31} {\rm~(eV)}^2 L {\rm~(km)}/\langle E_\nu \rangle
{\rm~(GeV)} = \pi/2$, i.e., $L/E_\nu = 353$~km/GeV for $\delta m^2_{31}
= 3.5\times10^{-3}$~eV$^2$. This has three important advantages: (i) the
$\nu_\mu \to \nu_\tau$ oscillation (which has only small matter effects)
is maximal, (ii) the $\nu_\mu \to \nu_e$ oscillation is nearly maximal,
{\it even when matter effects are taken into account}~\cite{bmw}, and
(iii) in the relevant limits that $\theta_{13}$ and $\delta
m^2_{21}/\delta m^2_{31}$ are small, the $\delta$ dependence is pure
$\sin\delta$, {\it even in the presence of matter}~\cite{bmw}. The
latter fact implies that there is no $\delta$--$\theta_{13}$ ambiguity
for a given sgn($\delta m^2_{31}$). There is a $\delta$--$(\pi-\delta)$
ambiguity, but it does not confuse a $CP$ violating ($CPV$) solution
with a $CP$ conserving ($CPC$) one. However, for small enough
$\theta_{13}$ and/or $L$, there is a ($\delta,\theta_{13}$)--sgn($\delta
m^2_{31}$) ambiguity, which sometimes can confuse $CPV$ and $CPC$
solutions; when combined with the $\delta$--$(\pi-\delta)$ ambiguity it
results in an overall four--fold ambiguity in parameters in these
cases~\cite{bmw}. Thus distinguishing the sign of $\delta m^2_{31}$ may
be essential for determining the existence of $CPV$.

We assume a narrow band beam (NBB) with flux $4\times10^{11}$/m$^2$/yr
at $L = 730$~km (and proportional to $1/L^2$), which would be about
${1/5}$ of the flux (to represent the flux loss in making a NBB) of
an upgraded NuMI ME beam with a 1.6~MW proton driver. The NBB has two
advantages: (i) the lack of a significant high--energy tail reduces
backgrounds, and (ii) nearly all of the neutrinos will be at the same
$L/E_\nu$, which is chosen near the peak of the oscillation. For simplicity, we
work in the monoenergetic approximation.

We assume an effective 70~kt-yr of data accumulation for detecting
$\nu_e$'s, which could be achieved by 2 years of running with a 70~kt liquid Argon detector~\cite{cline} at 50\% efficiency~\cite{superbeams}. For
$\nu_\tau$ detection we assume 3.3~kt-yr (2 years with a 5~kt detector
at 33\% efficiency).  For $\bar\nu$'s, we assume approximately 6--12
years of running (a factor of two longer to account for the lower
$\bar\nu$ cross section and another factor of 1.5--3 longer, depending
on $E_\nu$, to account for the reduced $\bar\nu$ flux in the beam). Thus
in the absence of matter and/or $CPV$ the number of $\nu$ and $\bar\nu$
events would be the same. We assume a $\nu_e$ background of $0.4\%$ of
the unoscillated CC signal, and a fractional uncertainty of the
background of 10\%.

We expect $\delta m^2_{21}$ to be measured to 10\% accuracy at
KamLAND~\cite{bmwood}, and $\delta m^2_{31}$ to be measured to about the
same accuracy by K2K, MINOS, and ICANOE, and OPERA. Since $E_\nu$ is
chosen to be at the peak of the leading oscillation, the choice of
$E_\nu$ depends critically on the value of $\delta m^2_{31}$; also, the
size of the $CPV$ and the potential for confusion between $\delta
m^2_{31} > 0$ and $\delta m^2_{31} < 0$ increases with increasing
$\delta m^2_{21}$. Our results for $\delta m^2_{31} > 0$ 
with $\theta_{23}=\pi/4$ are presented
in Table~\ref{tab:scenarios} for two values of $\delta m^2_{21} =
5\times10^{-5}$~eV$^2$ (the value preferred from recent 
analyses~\cite{solar1,solar2,solar3,solar4} of solar neutrino data) 
and $\delta m^2_{21} = 10^{-4}$~eV$^2$; the
corresponding results for $\delta m^2_{31} < 0$ are found by
interchanging $\langle N_e \rangle \leftrightarrow \langle \bar N_e
\rangle$ and $(\nu_\mu\to\nu_e) \leftrightarrow
(\bar\nu_\mu\to\bar\nu_e)$. For each value of $\delta m^2_{21}$ we show
results for three values of $\delta m^2_{31}$ that cover the range
inferred from Super--K atmospheric neutrino data. Given in the table are
(i) the numbers of $e$ and $\bar e$ events (for $\sin^22\theta_{13} = 0.01$ and
averaged over $\delta$), background $e$ events ($B_e$, assumed the same
for $e$ and $\bar e$), and $\tau$ events, (ii) the $\sin^22\theta_{13}$
reach at $3\sigma$ for $\nu_\mu \to \nu_e$ and $\bar\nu_\mu \to
\bar\nu_e$ appearance, and the minimum $\sin^22\theta_{13}$ for which
sgn($\delta m^2_{31}$) can be determined, and (iii) the smallest value
of the $CP$ phase $\delta$ that can be distinguished from $\delta =
0,\pi$ at the $3\sigma$ level for $\sin^22\theta_{13} = 0.01$ (not
accounting for a possible sgn($\delta m^2_{31}$) ambiguity). The
$\sin^22\theta_{13}$ reaches and $\delta$ sensitivity include the
effects of statistical and systematic experimental uncertainties. The
$e$ and $\bar e$ event rates approximately scale with
$\sin^22\theta_{13}$. Results for JHF--SK running for 5 years with
neutrinos only~\cite{jhfsk}, using a $2^\circ$ off axis beam,
are also shown in the table.

\begin{table}[t]
\squeezetable
\caption[]{Scenarios with $\delta m^2_{31} > 0$ (2 years $\nu$, 6--12
years $\bar\nu$); the last entry in the table shows the results for
JHF--SK~\cite{jhfsk} (5 years, $\nu$ only). $\theta_{23}=\pi/4$ is assumed.}
\label{tab:scenarios}
\begin{tabular}{|r|r|rr|rrrr|lll|c|}
$\delta m^2_{21}$ &$\delta m^2_{31}$ & $L$ & $E$
& $\langle N_e \rangle$ & $\langle \bar N_e \rangle$ & $B_e$ & $N_\tau$
& \multicolumn{2}{r}{$\sin^22\theta_{13}$ reach at $3\sigma$} & & $|\delta|$
($^\circ$) at $3\sigma$\\
(eV$^2$) &(eV$^2$) & (km) & (GeV)
& \multicolumn{2}{c}{$\sin^22\theta_{13} = 0.01$}
& & & $\nu_\mu \to \nu_e$ & $\bar\nu_\mu \to \bar\nu_e$
& sgn($\delta m^2_{31}$) & $\sin^22\theta_{13} = 0.01$ \\
\hline
$5\times10^{-5}$
& $2\times10^{-3}$
&  350 & 0.57 & 180 & 148 & 116 & -- & 0.0020 & 0.0025 & --     & 26\\
&&  730 & 1.18 &  95 &  63 &  56 & -- & 0.0026 & 0.0042 & 0.10   & 35\\
&& 1290 & 2.09 &  64 &  27 &  32 & -- & 0.0031 & 0.0082 & 0.036  & 49\\
&& 1770 & 2.86 &  53 &  15 &  23 & -- & 0.0033 & 0.014  & 0.020  & 67\\
&& 2900 & 4.70 &  39 &   4 &  14 & 10 & 0.0038 & 0.055  & 0.011  & --\\
\hline
& $3.5\times10^{-3}$
& 350 & 0.99 & 293 & 237 & 204 &  -- & 0.0024 & 0.0029 & --     & 39\\
&& 730 & 2.07 & 156 & 100 &  97 &  -- & 0.0026 & 0.0042 & 0.050  & 52\\
&&1290 & 3.65 & 106 &  42 &  55 &  14 & 0.0027 & 0.0073 & 0.015  & --\\
&&1770 & 5.01 &  88 &  22 &  40 &  36 & 0.0028 & 0.012  & 0.0091 & --\\
&&2900 & 8.22 &  67 &   5 &  25 &  51 & 0.0029 & 0.043  & 0.0057 & --\\
\hline
& $5\times10^{-3}$
& 350 & 1.41 & 412 & 331 & 289 &  -- & 0.0024 & 0.0030 & 0.098  & 54\\
&& 730 & 2.96 & 219 & 139 & 139 &  -- & 0.0025 & 0.0040 & 0.028  & 83\\
&&1290 & 5.21 & 150 &  57 &  79 &  77 & 0.0025 & 0.0066 & 0.0095 & --\\
&&1770 & 7.16 & 125 &  30 &  58 & 100 & 0.0025 & 0.011  & 0.0061 & --\\
&&2900 &11.74 &  95 &   7 &  35 & 102 & 0.0025 & 0.036  & 0.0041 & --\\
\hline\hline
$10^{-4}$
& $2\times10^{-3}$
&  350 & 0.57 & 233 & 201 & 116 & -- & 0      & 0      & --    & 14\\
&&  730 & 1.18 & 120 &  88 &  56 & -- & 0      & 0      & --    & 18\\
&& 1290 & 2.09 &  78 &  41 &  32 & -- & 0.0007 & 0.0019 & 0.10  & 24\\
&& 1770 & 2.86 &  62 &  24 &  23 & -- & 0.0014 & 0.0059 & 0.055 & 30\\
&& 2900 & 4.70 &  44 &   9 &  14 & 10 & 0.0025 & 0.036  & 0.023 & 51\\
\hline
& $3.5\times10^{-3}$
&  350 & 0.99 & 324 & 268 & 204 & -- & 0.0013 & 0.0016 & --    & 19\\
&&  730 & 2.07 & 170 & 114 &  97 & -- & 0.0017 & 0.0026 & --    & 24\\
&& 1290 & 3.65 & 114 &  50 &  55 & 14 & 0.0020 & 0.0052 & 0.040 & 32\\
&& 1770 & 5.01 &  94 &  28 &  40 & 36 & 0.0022 & 0.0092 & 0.021 & 40\\
&& 2900 & 8.22 &  69 &   8 &  25 & 51 & 0.0025 & 0.037  & 0.010 & 76\\
\hline
& $5\times10^{-3}$
& 350 & 1.41 & 433 & 353 & 289 &  -- & 0.0018 & 0.0023 & --     & 25\\
&& 730 & 2.96 & 229 & 149 & 139 &  -- & 0.0020 & 0.0032 & 0.081  & 31\\
&&1290 & 5.21 & 148 &  55 &  79 &  77 & 0.0021 & 0.0056 & 0.022  & 40\\
&&1770 & 7.16 & 129 &  34 &  58 & 100 & 0.0022 & 0.0092 & 0.012  & 50\\
&&2900 &11.74 &  96 &   9 &  35 & 102 & 0.0023 & 0.033  & 0.0063 & --\\
\hline\hline
-- & $3\times10^{-3}$
&  295 & 0.7 & 12 & -- & 22 & -- & 0.016 & -- & --    & --\\
\end{tabular}
\end{table}

In most cases the $\nu_\mu \to \nu_e$ appearance reach is about
0.002--0.003 for $\delta m^2_{21} = 5\times10^{-5}$~eV$^2$, and improves in the larger
$\delta m^2_{21}$ case (where even for $\sin^22\theta_{13}=0$ there is a
signal due to the subleading oscillation). The $\bar\nu_\mu \to
\bar\nu_e$ appearance reach is generally about 0.003 at small $L$,
decreasing to about 0.04--0.05 near $L=2900$~km, primarily due to the matter
suppression of antineutrinos for $\delta m^2_{31} > 0$. This matter
suppression and the $1/L^2$ dependence of the flux leads to decreased
$CPV$ sensitivity at larger $L$, especially for larger $\delta
m^2_{31}$.  However, larger $L$ does better at distinguishing
sgn($\delta m^2_{31}$) due to strong matter effects, and has higher
$\tau$ event rates because of the higher $E_\nu$, well above the $\tau$
production threshold at $E_\nu = 3.56$~GeV. Shorter $L$ values have
better $\delta$ sensitivity, except that there is potential confusion
with a different value of $\delta$ having the opposite sgn($\delta
m^2_{31}$), which in some cases could include a $CPV$/$CPC$ confusion;
also, $E_\nu$ is generally below the $\tau$ threshold.

If $\delta m^2_{21}$ is at the low end of its expected range, $CPV$ can
only be tested at shorter $L$, with the loss of the $\tau$ signal and
sgn($\delta m^2_{31}$) determination sensitivity, and potential
$CPV$/$CPC$ confusion due to sgn($\delta m^2_{31}$) (the four--fold
ambiguity mentioned above)~\cite{bmw}. Longer $L$ (such as $L=2900$~km)
could potentially do everything except for $CPV$, although if $\delta
m^2_{31}$ is too low $\tau$'s are not observable.  If $\delta m^2_{31}
\simeq 2\times10^{-3}$~eV$^2$ and a large $\tau$ signal is desired, then
the strategy outlined in this report will not work; $E_\nu$ must be
increased, which would force $L/E_\nu$ to be off the peak of the
oscillation.

For $L = 1290$ or 1770~km it is possible to simultaneously have
good $\sin^22\theta_{13}$ reach and sgn($\delta
m^2_{31}$) determination, and possibly sizeable $\tau$ rates and
 some $\delta$ sensitivity
if both $\delta m^2_{21}$ and $\delta m^2_{31}$ are at the high end of
their expected ranges (see Table~\ref{tab:scenarios}); 
$L = 1770$ km is probably preferred in
these cases due to its larger $\tau$ rate and better sgn($\delta
m^2_{31}$) determination.

We note that while a larger $\delta m^2_{21}$ in principle improves the
$CPV$ sensitivity, it also makes a sgn($\delta m^2_{31}$) ambiguity more
likely, leading to an overall four--fold ambiguity. Even if sgn($\delta
m^2_{31}$) is determined, measurements on the oscillation peak will leave
a two--fold ambiguity between $\delta$ and $\pi-\delta$. Measurements at
different $L$ and/or $E_\nu$ will be required to resolve these
ambiguities~\cite{bmw}.

\begin{acknowledgments}
We gratefully acknowledge helpful discussions with S. Geer and D. Harris.
This research was supported by the U.S.~Department of Energy
under Grants No.~DE-FG02-94ER40817 and No.~DE-FG02-01ER41155, by a DPF
Snowmass Fellowship and
 by the University of Wisconsin Research Committee with funds
granted by the Wisconsin Alumni Research Foundation.
\end{acknowledgments}

\end{document}